%% file: ChargeAccNoRad.tex
\documentclass[aps,pra,twocolumn,amssymb,amsmath,10pt]{revtex4-2}
\usepackage{amsmath}
\usepackage{amssymb}
\usepackage{epsfig}

\usepackage[colorlinks]{hyperref}

\hypersetup{
citecolor=magenta
}

\DeclareMathOperator{\sinc}{sinc}

\newcommand{\ket}[1]{\left | #1 \right \rangle}
\newcommand{\bra}[1]{\left \langle #1 \right |}

\newcommand{\beq}{\begin{equation}}
\newcommand{\eeq}{\end{equation}}
\newcommand{\ra}{\rangle}

\newcommand{\hp}{{\hat p}}

\begin{document}

\title{Charge acceleration without radiation}

\author{Yakir Aharonov$^{a,b,c}$, Daniel Collins$^{d}$ and Sandu Popescu$^{d}$ }

\affiliation{$^a$Schmid College of Science and Technology, Chapman University, Orange, California 92866, USA}
\affiliation{$^b$Institute for Quantum Studies, Chapman University, Orange, California 92866, USA}
\affiliation{$^c$School of Physics and Astronomy, Tel Aviv University, Tel Aviv 6997801, Israel}
\affiliation{$^{d}$H. H. Wills Physics Laboratory, University of Bristol, Tyndall Avenue, Bristol BS8 1TL}

\date{Nov 2025}
\begin{abstract}

The existence of electromagnetic radiation - radio-waves, microwaves, light, x-rays and so on - is one of the most important physical phenomena, and our ability to manipulate them is one of the most significant technological achievement of humankind. Underlying this ability is our understanding of how radiation is produced: whenever an electric charge is accelerated, it radiates. Or, at least, this is how it has been hitherto universally thought. Here we prove that quantum mechanically electric charges can be accelerated without radiating.  The physical setup leading to this behavior is relatively simple (once one knows what to do) but its reasons are deep: it relies on the fact that quantum mechanically particles can be accelerated even when no forces act on them, via the Aharonov-Bohm effect. As we argue, the effect presented here is just the tip of an iceberg - it implies the need to reconsider the basic understanding of radiation. Finally, it seems clear that the effect goes far beyond electromagnetism and applies to any kind of radiation. 

\end{abstract}

\maketitle

\section{Introduction}

Accelerated electric charges radiate. Arguably, this is one of the most important physical phenomena. Understanding the origin of radiation by J.C. Maxwell \cite{Maxwellemwaves} has been one of the greatest triumphs of physics.  In going from classical electromagnetism to quantum electrodynamics (QED), the same idea, i.e. that a charge undergoing a acceleration radiates, is generally presumed to be true. However, the theory is formulated in a completely different manner than the classical one. In quantum mechanics the notion of acceleration does not play a central role; it does not even enter explicitly in the basic equations. Given this, we raise the question whether the general presumption that accelerated charges radiate is true. The answer, as we show here, is that quantum mechanically it is possible to accelerate charges without them radiating.

The issue stems from the fundamental difference between classical and quantum dynamics. In classical physics forces are necessary to change how a system moves. In quantum mechanics the behaviour of systems can be affected even when particles move in force-free regions. Aharonov and Bohm discovered \cite{aharonovBohm} that changes in interference patterns in electron interference experiments may occur even though the electron never experiences any force. In subsequent work \cite{modularVariables, yakirModularMomentum}, Aharonov et al. explicitly showed that the {\it momentum distribution} of quantum particles can be changed without forces. This is dynamical non-locality \cite{dynamicalNonlocality} - for a particle whose state is composed of a superposition of wavepackets separated in space, the effect depends on the difference of potential between the remote locations of the different wavepackets, and it is non-zero even though the potential is constant (hence no force) at the location of each wavepacket. Of course, a change of momentum of a (non-relativistic) particle involves acceleration, hence quantum mechanically acceleration can occur without forces. We will show that when acceleration is produced in this non-local way no radiation occurs. 

\section{Main Experiment}

Describing electromagnetic radiation in the full quantum electrodynamics (QED) theory in anything but trivial situations is complicated, but for our purpose it is enough to consider a regime in which full QED treatment is not necessary. We will consider the electron moving with non-relativistic speed, and situations in which no electron-positron pairs may be created. The main facts that are relevant for us are that an electron moving with constant velocity is characterised by a Coulomb field in which the magnitude of the electric field ${\bf E}$ decreases proportionally to $1/R^2$, while any radiation it may emit when perturbed is characterised by an electric field that, at least in some directions, decreases only by $1/R$, where $R$ is the distance from the electron to the point where the field is observed \cite{thomson,feynmanI}.  Hence radiation can be observed and separated from the Coulomb field.

\begin{figure}[ht]
\def\svgwidth{9.0cm}
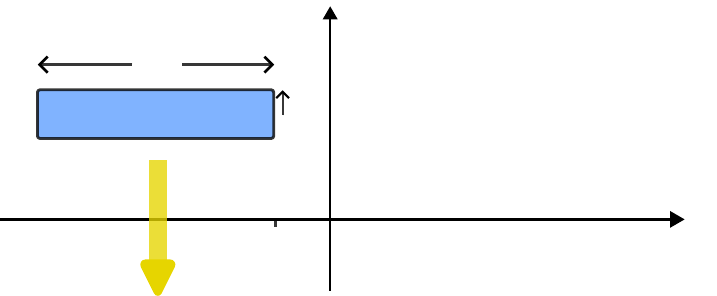
\caption{An electron starts in a wavepacket aligned above the x-axis, and moves downwards in the z-direction.}
\label{Fig:ChargeAccNoRadFig1}
\end{figure} 

Let us first consider an electron whose state $\Psi_L(x,y,z)$ is a cylindrical wavepacket of length $d$ and small radius $r$, with the axis along the $x$ axis from $x= -(d+a)$ to $x=-a$, with positive $a$, $y=0$ and $z=z_0>r$, as illustrated in Fig. \ref{Fig:ChargeAccNoRadFig1}. Let the wavepacket move in the direction of negative $z$. 

What will happen to the electron and its electromagnetic field during its time evolution? 

The answer is that there is no radiation.  This is because it is a free particle, and its momentum doesn't change - no acceleration, no radiation.  We can look at it in more detail to convince ourselves that although we are talking about a superposition of momentum eigenstates (since the electron is initially confined in a limited space region), there is indeed no radiation.  Imagine first an electron at rest, i.e. in a state of zero momentum.  Obviously then it will not radiate: it simply does not have any energy to give away.  If instead of being at rest, it starts in an eigenstate of well defined non-zero momentum, it will also not radiate, since this is equivalent to the original state at rest viewed in a boosted frame.  Finally, by linearity of quantum mechanics, since none of the terms in this superposition radiates, the superposition does not radiate.

\begin{figure}[ht]
\def\svgwidth{9.0cm}
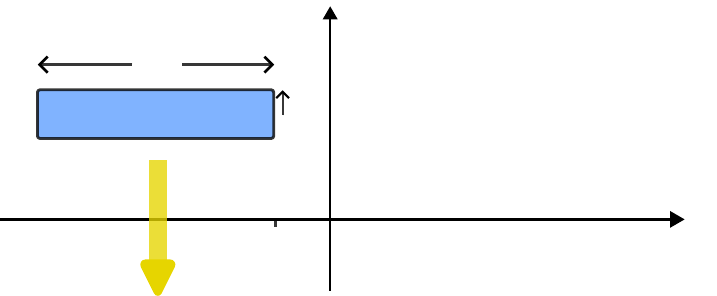
\caption{An electron moves past a solenoid with flux $\Phi$ which is oriented along the y-axis.  We use the singular gauge for the vector potential ${\bf A}$, taking it to be non-zero only along the half-plane with $z=0$ and $x>0$.}
\label{Fig:ChargeAccNoRadFig2}
\end{figure}

Consider now an infinitely long and thin solenoid containing a magnetic flux $\Phi$ oriented parallel to the $y$ axis and intersecting the $x-z$ plane at $z=0$, as illustrated in Fig. \ref{Fig:ChargeAccNoRadFig2}. 

Suppose that the wavepacket moves quickly enough so that the electron wavefunction, although spreading, does not enter the solenoid during the time it passes by it. What will happen now? 

The answer is that there will still be no radiation. The reason is that there is no electromagnetic field at the location of the electron, and the electron does not touch the solenoid nor encircle it, so the electron behaves as a free particle. Hence the wavepacket does not change its evolution versus what it would have been in the absence of the solenoid, in which case, as discussed, the electron does not radiate.

\begin{figure}[ht]
\def\svgwidth{9.0cm}
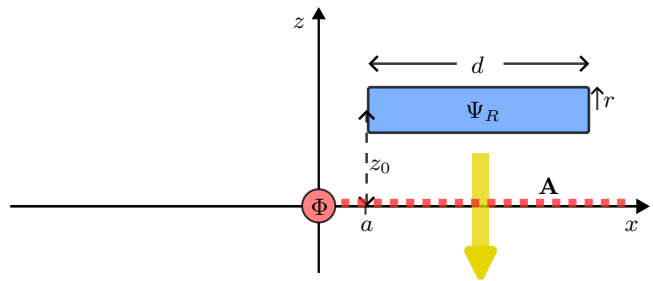
\caption{The electron moves past the right of the solenoid.  It crosses the singular line of the vector potential, acquiring an overall phase in our gauge.}
\label{Fig:ChargeAccNoRadFig3}
\end{figure}

Consider now that instead of being prepared in $\Psi_L$ the electron is prepared in the state $\Psi_R$, which is identical to $\Psi_L$ but which is placed from $x=a$ to $x=d+a$, i.e. on the other side of the $y-z$ plane to $\Psi_L$ (Fig. \ref{Fig:ChargeAccNoRadFig3}).  Now the electron passes on the right side of the solenoid and, similar to the previous case, there is no radiation.

\begin{figure}[ht]
\def\svgwidth{9.0cm}
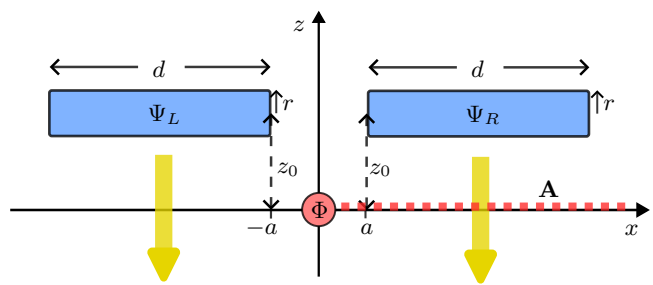
\caption{The electron, in the superposition $\frac{1}{\sqrt{2}}(\Psi_L + \Psi_R)$, moves past the solnoid.  The momentum distribution changes, without radiation.}
\label{Fig:ChargeAccNoRadFig4}
\end{figure}

Finally, consider that the electron was initially prepared in a superposition of the two wavepackets as in Fig. \ref{Fig:ChargeAccNoRadFig4},
\beq 
\Psi = \frac{1}{\sqrt 2}(\Psi_L+\Psi_R).
\eeq

Due to the linearity of quantum mechanics, the time evolution of each wavepacket is the same as when it is alone and not in the superposition. Hence none of the terms in the superposition radiates. Also, the quantum superposition of Coulomb fields of the two terms cannot result in radiation (via some possible interference), since each decays by $1/R^2$, and even if they interfere, the interference pattern will also decay as $1/R^2$, which is much faster than the $1/R$ decay of radiation.  Therefore, indeed, the superposition cannot radiate.

Crucially, however, there exists a significant effect of the solenoid on the superposition $\Psi$ of two wavepackets which passes it on either side - the so called Aharonov-Bohm effect \cite{aharonovBohm}. Indeed, as we noted above, since none of the individual wavepackets experiences the magnetic field, each of them, when alone, behaves in the same way as it would in the absence of the solenoid. Nevertheless, this does not rule out the possibility of each of them accumulating an overall phase. Since the solenoid contains a magnetic flux $\Phi$, the integral of the vector potential $\bf A$ on a closed loop around the solenoid should give the magnetic flux:
\beq 
\oint {\bf A}d{\bf l} = \Phi.
\eeq

For simplicity, let us use the singular gauge in which the vector potential is non-zero only on the $x>0$ part of the $z=0$ plane, where the vector potential $\bf A$ is oriented perpendicular to this plane (i.e. $A_x=A_y=0$) and has magnitude $A=A_z=\delta(z)\Phi$.  All the physically observable implications will, however, be gauge invariant.  In particular, for later use, note that, as opposed to momentum, velocity is a gauge invariant quantity. 

In the presence of the vector potential each wavepacket accumulates an overall phase depending on its path $\Gamma$. The electron in the $\Psi_L$ wavepacket doesn't encounter any vector potential hence it accumulates a zero phase
\beq 
\varphi_1 = q \int_{\Gamma_1} {\bf A} d{\bf l} = 0
\eeq 
while the $\Psi_R$ wavepacket crosses the vector-potential line and accumulates an overall phase of 
\beq 
\varphi_2 = q \int_{\Gamma_2} {\bf A} d{\bf l} = q \Phi
\eeq
where $q$ is the charge of the electron, and we take $\hbar = 1$.  Consequently, the state of the electron becomes
\beq 
\Psi_{\alpha} = \frac{1}{\sqrt{2}} ( \Psi_L+e^{i\alpha}\Psi_R )
\eeq
with 
\beq 
\alpha = q \Phi
\eeq
The individual phases $\varphi_1$ and $\varphi_2$ are, of course, gauge dependent, but the relative phase $\alpha$ is gauge independent.

\begin{figure}[ht]
\def\svgwidth{9.0cm}
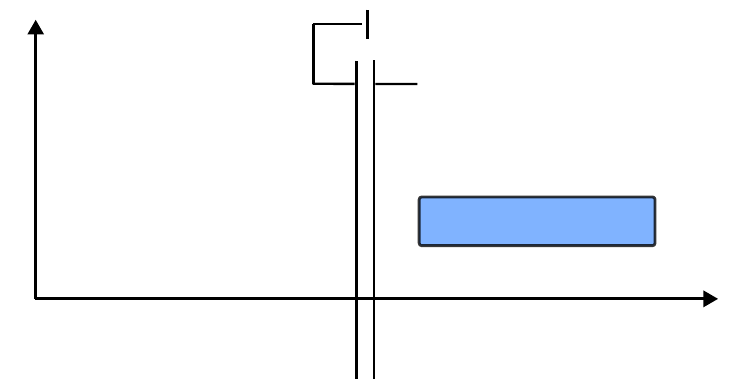
\caption{An infinite parallel plate capacitor, $C$, is placed between the left and right wavepackets of an electron in the superposition $\frac{1}{\sqrt{2}}(\Psi_L + \Psi_R)$.  The capacitor is charged and discharged, which results in a relative phase between the two wavepackets but no radiation, similar to the action of the solenoid.}
\label{Fig:ChargeAccNoRadFig6}
\end{figure}

An alternative way to obtain the same effect is by using a capacitor with infinite parallel plates in between the two wavepackets (Fig. \ref{Fig:ChargeAccNoRadFig6}), or temporarily enclosing each wave packet in a Faraday cage, and applying a time dependant electric potential difference in between the capacitor plates or the two Faraday cages. For details see Appendix \ref{appendixCapacitor}.

Now comes the main point. The effect of the relative phase is to modify the distribution of the $x$ component of the momentum of the electron.  In our particular gauge (where both initially and finally $v = p/m$) this implies a change in the distribution of the $x$ component of the velocity, therefore acceleration. 

In our example, the two wavepackets are identical, up to a shift in the $x$ direction. For simplicity, let them be of the form 
\beq 
\phi(x,y,z)=\Theta(x|d) \chi(y,z)
\eeq 
where $\Theta$ is the normalized top-hat function of length $d$, 
\beq
\Theta(x|d) = \begin{cases}
    \frac{1}{\sqrt d} & 0 < x \le d\\
    0 & otherwise,\\
\end{cases}
\eeq
and whose momentum representation is 
\beq {\tilde \Theta}(p_x|d)= \sqrt{\frac{d}{2 \pi}}e^{-i p_x d/2}\sinc(p_x \frac{d}{2}),\eeq 
where $\sinc(x) = \sin(x)/x$.
Had the particle been prepared in a single wavepacket, the probability (density) distribution of $p_x$ would be
\beq
{\cal P}_{single}(p_x)=\frac{d}{2\pi}\sinc^2(p_x \frac{d}{2}).
\eeq
Instead we have a superposition of two wavepackets,
\beq
\label{initialStateTwoWavepackets}
\Psi(x,y,z) = \frac{1}{\sqrt{2}} \Big(\Theta(x + d + a|d)+\Theta(x - a|d)\Big) \chi(y,z).
\eeq

The $x$-momentum (density) distribution ${\cal P}(p_x)$ of the initial state $\Psi$ (obtained in Appendix \ref{appendixTwoWavepackets} by making the Fourier transform of the $x$-dependent part of the wavefunction) is 
\beq 
{\cal P}(p_x) = \frac{d}{\pi} \cos^2 (p_x \frac{D}{2} ) \sinc^2(p_x \frac{d}{2}),
\eeq
where $D = d + 2a$.
Note that the effect of the superposition is to modulate the momentum probability distribution of the individual wavepackets by the factor $\cos^2(p_xD/2)$.

The final superposition, after the particle has passed the solenoids, is
\beq
\Psi_{\alpha}(x,y,z) = \frac{1}{\sqrt{2}} \Big(\Theta(x + d + a|d)+ e^{i \alpha} \Theta(x - a|d)\Big) \chi(y,z).
\eeq
Correspondingly, the final $x$-momentum distribution ${\cal P'}(p_x)$ is
\beq 
{\cal P'}(p_x) = \frac{d}{\pi} \cos^2( \frac{p_x D - \alpha}{2}) \sinc^2(p_x \frac{d}{2}).
\eeq

\begin{figure}[ht]
\includegraphics[width=8.6cm]{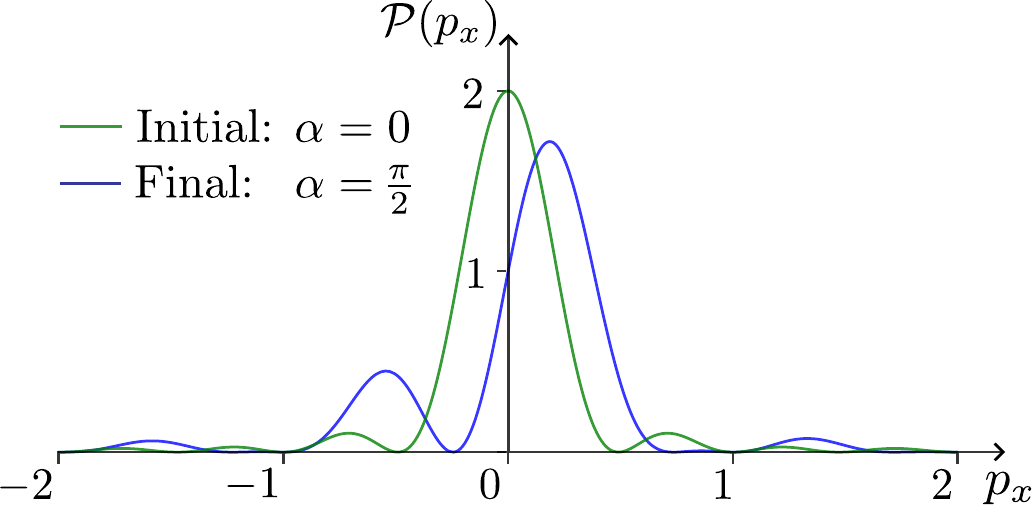}
\caption{The $x$-momentum probability density, initially ${\cal P}(p_x)$ in green, and after passing the solenoids ${\cal P'}(p_x)$ in blue, for $d=2 \pi$, $a=0.01 d$, $D = d + 2a$ and $\alpha = \pi/2$.  The peak of the final distribution is to the right of the initial one.}
\label{Fig:ChargeAccNoRadGraph1}
\end{figure} 

As one can see, except the cases when $\alpha=2n\pi$, the final distribution of $p_x$ momentum, ${\cal P}'(p_x)$ is different from the initial one, ${\cal P}(p_x)$. For example,for $\alpha=\pi/2$ the peak of the momentum distribution is now at approximately $p_x=1.2/d$, for small $a$ while initially it was at $p_x=0$, as illustrated in Fig. \ref{Fig:ChargeAccNoRadGraph1}. In case of $\alpha=\pi$ (not illustrated in Fig. \ref{Fig:ChargeAccNoRadGraph1}) there is zero probability of finding $p_x=0$ in the final state. Clearly, in all the cases when the momentum distribution has changed, the electron was accelerated. Yet, there was no radiation.

One may wonder what happens after the relative phase was introduced. The relative phase affects the probability distribution of momentum.  However, as long as the two wavepackets are separated, it has no effect on the spatial distribution of the electron.  The effect of the relative phase on the spatial distribution can only be seen later when the two wavepackets overlap: their interference depends on the relative phase.  One may wonder whether or not at this time the electron starts radiating.  In other words, whether radiation was not simply just delayed for the time when the relative phase makes an impact on the spatial structure of the electron.  The answer is no: there will be no radiation at any time.  The simple, basic argument is that after the relative phase was accumulated, the particle evolves freely, and free particles do not radiate.

As a more extreme case we will now show a situation in which an electron whose initial momentum is $p=0$, within an approximation as good as we want, could be accelerated to an arbitrary given momentum $p_0$, with probability as close to 1 as we want, without radiation. To achieve this, we will arrange an initial situation in which the electron is prepared in a superposition of many wavepackets, and then change all their relative phases, by an array of solenoids. By an argument identical to that above, there will be no radiation.

Like in the simple example above, the main effect relates to acceleration along the $x$ axis, so we will consider again the electron wavefunction to be a direct product, $\Psi_0(x)\chi(y,z)$. Let us now focus on the $x$ coordinate.

\begin{figure}[ht]
\def\svgwidth{10.0cm}
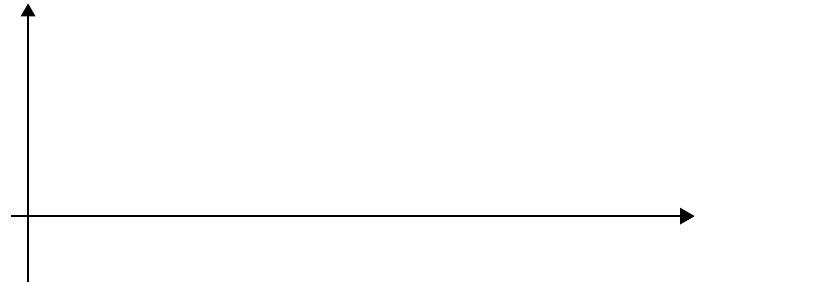
\caption{The electron, in a state of N wavepackets with gaps in-between, passes through an array of solenoids without touching them.  Each solenoid has a singular line of vector potential in the positive $x$-direction giving a phase $\alpha$, thus the $n^{th}$ wavepacket acquires a phase $n \alpha$.  The electron's $x$-momentum distribution changes from being close to $0$ to close to $p_0 = \alpha / l$.}
\label{Fig:ChargeAccNoRadFig5}
\end{figure}

Consider the initial wavefunction $\Psi_0(x)$ which is a superposition of $N$ top-hat wavepackets of length $d$ separated by a distance $\epsilon$ with $\epsilon/l=\xi$:
\beq
\Psi_0(x)=\frac{1}{\sqrt{N}}\sum_{n=0}^{N-1}\Theta(x-nl|d)
\eeq
where $l=d+\epsilon$ is the periodicity of the superposition and $L=Nl$ is the total length of a similar wavefunction without gaps (including a gap after the last non-zero piece, see Fig. (\ref{Fig:ChargeAccNoRadFig5}).

In the limit of small gap fraction, $\xi\ll 1$, and large $L$, this function has a momentum distribution as close to $p=0$ as we want.  We can easily understand the large $L$ small $\xi$ limit: Taking the gap fraction smaller, we approximate a simple top-hat function of length $L$ whose average momentum is $p=0$. To minimize the spread around $p=0$, we shall take $L$ to be large. Mathematically, to see that having a small gap fraction makes $\Psi_0$ approximate the top-hat function $\Theta (x|L)$ is trivial - one can see that even just by looking at their plots. Alternatively, by computing the scalar product we find it to be $\bra{\Psi_0} \Theta(x|L)\ra=1-O(\xi)$.  This also means that the momentum probability density distribution is up to order $\xi$ close to that of the top-hat function of length $L$, which, in its turn, is peaked closer and closer to $p=0$ when $L$ increases:  The momentum probability density of $\Psi_0$ is therefore very close to that of $\Theta(x|L)$, i.e. 
\beq 
\label{momentumDistributionZero}
{\cal P}(p; \Psi_0)\approx{\cal P}(p;\Theta)=\frac{L}{2 \pi} \sinc^2(p\frac{L}{2}),
\eeq 
for which by choosing $L$ of the order $1/\delta_p$ with $\delta_p$ as small as we want, we can make the probability of finding the momentum within the interval $ [-\delta_p, \delta_p]$ as close to 1 as we want. 

Next we accelerate the particle to our target momentum $p_0$, with probability as close to $1$ as desired, by passing it through an array of solenoids, placed so that they line up with the gaps in the wavefunction (Fig. \ref{Fig:ChargeAccNoRadFig5}).  Each of these solenoids gives the same phase, $\alpha$, so the final state is given by 
\beq
\Psi_{p_0}(x) = \frac{1}{\sqrt{N}}\sum_{n=0}^{N-1} e^{i n \alpha} \Theta(x-nl|d),
\eeq
where $\alpha = p_0 l$, which is the phase difference an eigenstate of momentum $p_0$ will acquire over a length $l$.  The cumulative effect of adding the same relative phase in between each subsequent wavepacket is that in between any two distant points along the $x$ axis there is a total relative phase that increases proportional to the distance, similar to the behaviour of a momentum eigenstate.  The difference is that in our case the phase does not increase continuously, but in a staircase fashion. Yet, if the steps are narrow (i.e. the wave is split into many very small wavepackets), this is a very good approximation.  

In more detail, we have already established that we need $L$ large and $\xi$ small, to ensure the probability of the momentum of the initial state being close to $0$ is as close to certainty as we want.  To ensure the probability distribution of the momentum of the final state is close to $p_0$ we only need one additional requirement: $l \ll \lambda=2 \pi/p_0$ where $\lambda$ is the wavelength of the desired final momentum $p_0$. The reason is that in a genuine momentum eigenstate the phase increases continuously with $x$, while in our case it increases stepwise, with step-length $l$. We simply need to make the steps smaller, specifically, smaller relative to $\lambda$.

\begin{figure}[ht]
\includegraphics[width=8.6cm]{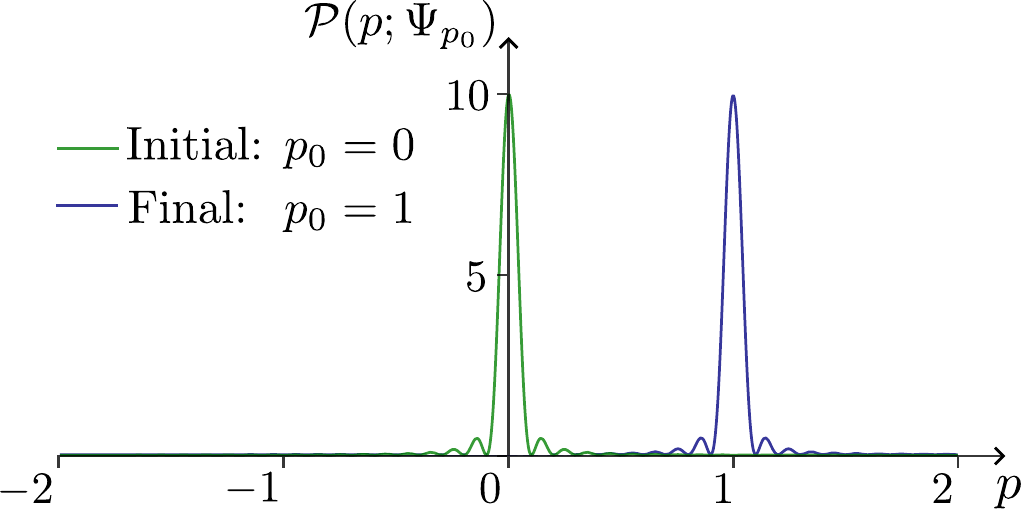}
\caption{The momentum probability density,  ${\cal P}(p;\Psi_0)$ in green, and ${\cal P}(p; \Psi_{p_0})$ in blue, for $L=20\pi$, $\xi \ll 1$, $l \ll \lambda = 2 \pi / p_0$ and $p_0 = 1$.  The peak of the final distribution is at $p_0$. The deviations of these probability distributions from $\sinc^2(pL/2)$ and $\sinc^2((p-p_0)L/2))$
respectively are too small to be visible at the scale of the figure. yet they are essential for determining the averages of the powers of momentum, which are strongly influenced by perturbations of very small probability, but large values of $p$. In particular, despite the visual appearance, the average value of $p$ is exactly the same for the two functions.\label{Fig:ChargeAccNoRadGraph2}}
\end{figure}

In the above limit, the final state $\Psi_{p_0}$ is as close as desired to that of the boosted top-hat function $e^{i p_0 x} \Theta(x;L)$ i.e. the scalar product $\bra{\Psi_{p_0}} e^{i p_0 x} \Theta(x|L) \ra = 1-O(\xi, l/\lambda)$ (see Appendix \ref{appendixStepFunctionCloseness}).  The probability density of $\Psi_{p_0}$ is then very close to that of $e^{i p_0 x} \Theta(x|L)$, i.e. 
\beq 
\label{momentumDistributionPzero}
{\cal P}(p; \Psi_{p_0}) \approx {\cal P}(p;e^{ip_0x} \Theta) = \frac{L}{2 \pi} \sinc^2((p-p_0)\frac{L}{2}),
\eeq 
which is strongly peaked around $p_0$ when $L$ is large (Fig. \ref{Fig:ChargeAccNoRadGraph2}).  Consequently, in the final state $\Psi_{p_0}$, the probability of finding the momentum within the interval $ [p_0-\delta_p, p_0 + \delta_p]$ can be made as close to $1$ as we like, by taking $L$ large, $\xi$ small, and $l \ll \lambda$.  

Hence we have achieved our goal of, roughly speaking, accelerating a charged particle from the initial state which has momentum $p=0$, to a final state which has momentum $p=p_0$, without radiation.

Let us look in more detail at the characteristics of this problem.  

How is it possible that we could accelerate a charged particle without radiation?  As we mentioned in our introduction, this stems from the key difference between classical and quantum mechanical dynamics. In classical mechanics, accelerating particles requires forces.  In quantum mechanics, one can accelerate particles without applying a force on them.  The Heisenberg equations of motion are non-local \cite{modularVariables,dynamicalNonlocality}.  This was first discovered in the context of the Aharonov-Bohm effect, but it is a universal feature whenever we have potentials or vector potentials. 

Looking at the wavefunctions  $\Psi_0$ and $\Psi_{p_0}$, we see that the peak changed, from $0$ to $p_0$, and that this peak occurs with probability almost one.  Although the wavefunctions $\Psi_0$ and $\Psi_{p_0}$ are very different from each other, with $\Psi_{0}$ very close to the eigenstate of momentum $0$, and the state $\Psi_{p_0}$ very close to the eigenstate of momentum $p_0$, the average values of momentum in $\Psi_0$ and $\Psi_{p_0}$ are the same (see Appendix \ref{appendixMomentsUnchanged}).  In fact, the averages of any power of momentum, i.e. all the moments of the probability distributions, are the same in both these states: $\bra{\Psi_0} \hp^n \ket{\Psi_0} = \bra{\Psi_{p_0}} \hp^n \ket{\Psi_{p_0}}$ for any integer $n$.  This is very strange, since the momentum has changed with probability almost $1$ from being close to $0$ to being close to $p_0$.  Also, the average value of momentum on the right-hand side of equations \eqref{momentumDistributionZero} and \eqref{momentumDistributionPzero}, which approximate our initial and final states with probability almost $1$, did change, from $p=0$ to $p=p_0$.  This is due to other, large, changes in momentum which occur with low probability, which are enough to keep all the averages unchanged.  

Nevertheless, we know that the probability distributions themselves did change, so we are faced with the unusual situation that our probability distributions changed without any of their moments changing.  What has changed are the modular variables \cite{modularVariables} of the form $\bra{\Psi} \hp \bmod (2 \pi/b) \ket{\Psi}$, or equivalently $\bra{\Psi} e^{-i \hp b} \ket{\Psi}$, for $\epsilon < b < L - \epsilon$.  These encode the non-local character of quantum mechanics.  The relevant operators, $e^{-i \hp b}$, are in fact the shift operators, which translate the position in space by a distance $b$ (in contrast the powers of momentum, $(-i \partial / \partial x)^n$, only produce localised changes to the wavefunction).  The particular values of $b$ for which the modular momentum changes are those where the shift operator moves our initial wavefunction enough so that at least one of the individual wavepackets is now superposed over another, from which it was initially separated, and allows interference which exposes the relative phase.  This behaviour is the main characteristic of the dynamical non-locality which underlies the AB effect.

\section{Discussion}

What characterised our examples is that the fields were localised in regions where the particle is not present.  Yet they have a non-local effect, by adding phases between different wave packets.  In this situation, there has been no radiation whatsoever despite accelerating the particle.  But what happens if the particle does enter a region where the field is present?  The main effect remains: what is important is that whenever we have a localised field this may have both local and non-local effects.  

For example, suppose we would have started with a top-hat function $\Theta(x)$ without any holes and let the solenoids go through it.  We can decompose this wavefunction into a superposition of two: one wavefunction with holes, and another composed of tiny wavepackets, one in each hole.  Due to the linearity of quantum mechanics, the first term will evolve as in our example above: that is the particle will be accelerated and there will be no radiation whatsoever.  If we take the size of the holes to be very small, the magnitude of this first term is as close to $1$ as we want.  At the same time, the second term describes the wavefunction that enters the solenoid, and there it will be accelerated.  This will cause radiation, however the magnitude of this term, hence the probability that we will ever observe it, is close to zero.  This second term, which does experience locally the fields, is accelerated via a force, and this {\it does} change the average value of momentum. 

Importantly, and this is the essence of the entire phenomena, what happens with the piece of the wavefunction which enters the solenoid and gets accelerated by the direct application of the force, is independent from the acceleration undergone by the piece of the wavefunction outside the solenoid.  Indeed, all that matters for the piece outside the solenoid is the relative phase $e^{i \alpha} = e^{i q \Phi}$ between the wavepackets.  Increasing or decreasing the magnetic field inside the solenoid such that the total flux $\Phi$ increases/decreases by $2 \pi / q$, changes the force on the piece inside the solenoid, affects the radiation (which is only produced by this piece), and accounts for the entire change of the average momentum.  Yet it does not affect the piece outside, which accounts, with probability as close to $1$ as we like, for the acceleration experienced by the particle, which is, as discussed, radiation free.

\section{Conclusions}

To summarise, any interaction has two aspects: a local one and a non-local one.  The presence of a field has a local effect on the part of the wavefunction which enters its area, and a non-local effect on the rest of the wavefunction which surrounds this area.  The local one does produce radiation, and does change the average value of the momentum.  The non-local one does not produce radiation, does not change the average value of the momentum, however it does accelerate the particle. 
When the field is over an extended region, it is difficult to see these two effects separately.  Yet we postulate that these two effects are always present (indeed if in the example above, we would zoom in to what happens inside the solenoid to the tiny pieces which enter, we should be able to see even there local and non-local parts).

We conclude that the issue of radiation in quantum mechanics is far more subtle than is usually appreciated, and that the standard way of thinking of radiation, inspired by classical mechanics, has to be changed.  Contrary to the classical case, we {\it can} accelerate a particle without producing radiation. 

\section{Acknowledgements}
Sandu Popescu and Daniel Collins are supported by the European Research Council Advanced Grant FLQuant.

\bibliography{ChargeAccNoRad}

\appendix

\section{Acceleration using a capacitor}
\label{appendixCapacitor}

In most of this paper we use solenoids to accelerate a charged particle without radiation.  However we may also use, for example, capacitors, as illustrated in Fig. \ref{Fig:ChargeAccNoRadFig6}. 
Here we explain how that leads to the same effect as with the solenoids. 

We start with an electron which is in a superposition of two wavepackets, and place a capacitor between the two.  The capacitor is initially discharged, we then charge and discharge it again within a short time $T$ during which the wavepackets, though spreading, do not touch or enter the capacitor. Let $V(t)$, with $V(0)=V(T)=0$, be the electric potential difference between the capacitor plates. The electric and magnetic fields outside the capacitor are zero at all times, hence, if the electron is prepared in a single wavepacket (to the left or the right of the capacitor) it behaves as it were  a free particle and therefore there is no radiation. Then, by the linearity of quantum mechanics, even in a superposition there will be no radiation. Yet a relative phase
\beq
\label{capacitorPhase}
\alpha = q \int V(t) dt 
\eeq
will accumulate between the two wavepackets. This is easiest to see in the Coulomb gauge where the potential outside the capacitor is constant in space but time dependant.  The potential on the left side of the capacitor is $-V(t)/2$, and the potential on the right side is $+V(t)/2$, hence the wavepacket on the left accumulates a phase $- q \int V(t)dt/2$, the wavepacket on the right accumulates a phase $q \int V(t)dt/2$, leading to the phase difference in Eq. \eqref{capacitorPhase}.

Therefore we have applied a relative phase between the two wavepackets, just as we did with the solenoids, without causing radiation.  A similar situation is if we place each wavepacket in a Faraday cage, and apply a time dependent electric potential difference between the two cages.

\section{Momentum distribution for superposition of two wavepackets}
\label{appendixTwoWavepackets}

We shall calculate the momentum distribution for the superposition of two wavepackets with relative phase $\alpha$ (in one dimension):
\beq
\Psi(x) = \frac{1}{\sqrt{2}} \left(\Theta(x+d+a|d) + e^{i \alpha} \Theta(x-a|d)\right),
\eeq
where $\Theta$ is a normalized top-hat function of length $L$, 
\beq
\Theta(x) = \begin{cases}
    \frac{1}{\sqrt{d}} & 0 < x \le d\\
    0 & otherwise.\\
\end{cases}
\eeq

We Fourier transform this into momentum space, which gives
\beq
\begin{split}
\tilde{\Psi}(p) &= \frac{1}{\sqrt{2 \pi}} \int \Psi(x) e^{-i p x} dx \\
&= \frac{1}{\sqrt{4 \pi}} \int \left(\Theta(x + d + a|d) + e^{i \alpha} \Theta(x-a|d)\right) e^{-i p x} dx \\
&= \frac{1}{\sqrt{4 \pi}} \int \left( e^{i p (d+a)} \Theta(x|d) + e^{i \alpha -i p a} \Theta(x|d)\right) e^{-i p x} dx \\
&= \frac{1}{\sqrt{2}} \left( e^{ip(d+a)} + e^{-i (p a - \alpha)} \right) \tilde{\Theta}(p|d),
\end{split}
\eeq
where
\beq
\begin{split}
\tilde{\Theta}(p|d) &= \frac{1}{\sqrt{2 \pi}} \int \Theta(x|d) e^{-i p x} dx \\
&= \frac{1}{\sqrt{2 \pi d}} \frac{e^{-i p d} - 1}{-ip}.
\end{split}
\eeq
By rearranging the formula and taking out overall phases, we can rewrite this in the more customary form 
\beq
\tilde{\Psi}(p) = \sqrt{\frac{d}{\pi}} e^{i (p d + \alpha)/2} \cos(\frac{pD - \alpha}{2}) \sinc(p \frac{d}{2}),
\eeq
where $\sinc(x) = \sin(x)/x$ and $D=d+2a$.  The momentum distribution is then
\beq 
{\cal P'}(p) = \frac{d}{\pi} \cos^2( \frac{pD - \alpha}{2}) \sinc^2(p \frac{d}{2}).
\eeq

\section{The final state is close to a boosted top-hat}
\label{appendixStepFunctionCloseness}

Here we show that the scalar product of the final state of $N$ wavepackets,
\beq
\Psi_{p_0}(x) = \frac{1}{\sqrt{N}}\sum_{n=0}^{N-1} e^{i n \alpha} \Theta(x-nl|d),
\eeq
with the boosted top-hat function,
\beq
\Phi_{p_0}(x) = e^{i p_0 x} \Theta(x|L),
\eeq
where $\alpha = p_0 L/N = p_0 l$, can be made as close to $1$ as desired by taking $l \ll 2 \pi/p_0$.  Their scalar product is:
\beq
\begin{split}
\bra{\Phi_{p_0}} \Psi_{p_0} \ra &= \frac{1}{\sqrt{N L d}} \sum_{n=0}^{N-1} \int_{nl}^{nl+d} e^{-ip_0 x}  e^{i n \alpha} dx \\
&= \frac{1}{\sqrt{NLd}} \sum_{n=0}^{N-1} e^{i n \alpha} \frac{e^{-i p_0 (nl+d)} - e^{-i p_0 nl}}{-i p_0}\\
&= \frac{1}{\sqrt{NLd}} \frac{e^{-i p_0 d} - 1}{-i p_0} \sum_{n=0}^{N-1} e^{i n \alpha} e^{-i p_0 nl} \\
&= \sqrt{\frac{N}{Ld}} \frac{e^{-i p_0 d} - 1}{-i p_0}.
\end{split}
\eeq
Since we approximate $p_0 x$ by a step function with each step of length $d$, we want the phase accumulated in that step, $p_0 d$, to be small.  We take $l \ll 2 \pi/p_0$, which implies $d \ll 2 \pi/p_0$ (since $d<l$).  Then the scalar product is approximately
\beq
\frac{1}{\sqrt{ld}} \frac{(1 -i p_0 d) - 1}{-i p_0} = \sqrt{\frac{d}{l}} = \sqrt{1- \xi}.
\eeq
Since we have already taken the gap fraction, $\xi$, to be small, this scalar product can be made as close to $1$ as desired.

\section{Moments of momentum do not change under the AB effect}
\label{appendixMomentsUnchanged}

We shall show that if one starts with a superposition of two wavepackets in one dimension,
\beq
\Psi = \frac{1}{\sqrt 2}(\Psi_L+\Psi_R),
\eeq
where $\Psi_L$ does not overlap in space with $\Psi_R$, and then applies a relative phase $\alpha$ to get
\beq
\Psi_{\alpha} = \frac{1}{\sqrt 2}(\Psi_L+ e^{ i \alpha} \Psi_R),
\eeq
then the moments $\bra{\Psi_{\alpha}} \hp^n \ket{\Psi_{\alpha}}$ are unchanged for all integer $n$.  

We demonstrate this by computing the moments, and showing that they do not depend upon $\alpha$.  Explicitly: 
\beq
\begin{split}
&\bra{\Psi_{\alpha}} \hp^n \ket{\Psi_{\alpha}} \\
&= \int \Psi^*_{\alpha}(x) \left(-i \frac{\partial}{\partial x} \right)^n \Psi_{\alpha}(x) dx \\
&= \frac{1}{2} \int (\Psi^*_L+ e^{-i \alpha} \Psi^*_R) \left(-i \frac{\partial}{\partial x} \right)^n (\Psi_L+ e^{i \alpha} \Psi_R) dx \\
&= \frac{1}{2} (\bra{\Psi_L} \hp^n \ket{\Psi_L} + \bra{\Psi_R} \hp^n \ket{\Psi_R}),
\end{split}
\eeq
where we used the fact that the cross terms between $\Psi_L$ and $\Psi_R$, such as $\bra{\Psi_L} \left(-i \partial/\partial x \right)^n \ket{\Psi_R}$, evaluate to $0$ because they have no spatial overlap, since $-i\partial/\partial x$ does not slide $\psi(x)$ to a different spatial location, e.g. to $\psi(x+a)$.  

Hence the moments of the momentum distribution do not depend upon $\alpha$, and so do not change when we apply a non-local AB phase.  This argument applies in a similar way if we have many spatially separated wavepackets.


\end{document}

%% file: ChargeAccNoRadFig1.pdf_tex
\begingroup%
  \makeatletter%
  \providecommand\color[2][]{%
    \errmessage{(Inkscape) Color is used for the text in Inkscape, but the package 'color.sty' is not loaded}%
    \renewcommand\color[2][]{}%
  }%
  \providecommand\transparent[1]{%
    \errmessage{(Inkscape) Transparency is used (non-zero) for the text in Inkscape, but the package 'transparent.sty' is not loaded}%
    \renewcommand\transparent[1]{}%
  }%
  \providecommand\rotatebox[2]{#2}%
  \newcommand*\fsize{\dimexpr\f@size pt\relax}%
  \newcommand*\lineheight[1]{\fontsize{\fsize}{#1\fsize}\selectfont}%
  \ifx\svgwidth\undefined%
    \setlength{\unitlength}{343.09264656bp}%
    \ifx\svgscale\undefined%
      \relax%
    \else%
      \setlength{\unitlength}{\unitlength * \real{\svgscale}}%
    \fi%
  \else%
    \setlength{\unitlength}{\svgwidth}%
  \fi%
  \global\let\svgwidth\undefined%
  \global\let\svgscale\undefined%
  \makeatother%
  \begin{picture}(1,0.41214266)%
    \lineheight{1}%
    \setlength\tabcolsep{0pt}%
    \put(0,0){\includegraphics[width=\unitlength,page=1]{ChargeAccNoRadFig1.pdf}}%
    \put(0.20621229,0.30863266){\color[rgb]{0,0,0}\makebox(0,0)[lt]{\lineheight{1.25}\smash{\begin{tabular}[t]{l}$d$\end{tabular}}}}%
    \put(0.91798653,0.07010644){\color[rgb]{0,0,0}\makebox(0,0)[lt]{\lineheight{1.25}\smash{\begin{tabular}[t]{l}$x$\end{tabular}}}}%
    \put(0.34598066,0.06683265){\color[rgb]{0,0,0}\makebox(0,0)[lt]{\lineheight{1.25}\smash{\begin{tabular}[t]{l}$-a$\end{tabular}}}}%
    \put(0.19239251,0.24208746){\color[rgb]{0,0,0}\makebox(0,0)[lt]{\lineheight{1.25}\smash{\begin{tabular}[t]{l}$\Psi_L$\end{tabular}}}}%
    \put(0.42195941,0.37394395){\color[rgb]{0,0,0}\makebox(0,0)[lt]{\lineheight{1.25}\smash{\begin{tabular}[t]{l}$z$\end{tabular}}}}%
    \put(0.4037634,0.25906864){\color[rgb]{0,0,0}\makebox(0,0)[lt]{\lineheight{1.25}\smash{\begin{tabular}[t]{l}$r$\end{tabular}}}}%
    \put(0.39070315,0.16963286){\color[rgb]{0,0,0}\makebox(0,0)[lt]{\lineheight{1.25}\smash{\begin{tabular}[t]{l}$z_0$\end{tabular}}}}%
    \put(0,0){\includegraphics[width=\unitlength,page=2]{ChargeAccNoRadFig1.pdf}}%
  \end{picture}%
\endgroup%

%% file: ChargeAccNoRadFig2.pdf_tex
\begingroup%
  \makeatletter%
  \providecommand\color[2][]{%
    \errmessage{(Inkscape) Color is used for the text in Inkscape, but the package 'color.sty' is not loaded}%
    \renewcommand\color[2][]{}%
  }%
  \providecommand\transparent[1]{%
    \errmessage{(Inkscape) Transparency is used (non-zero) for the text in Inkscape, but the package 'transparent.sty' is not loaded}%
    \renewcommand\transparent[1]{}%
  }%
  \providecommand\rotatebox[2]{#2}%
  \newcommand*\fsize{\dimexpr\f@size pt\relax}%
  \newcommand*\lineheight[1]{\fontsize{\fsize}{#1\fsize}\selectfont}%
  \ifx\svgwidth\undefined%
    \setlength{\unitlength}{343.09264656bp}%
    \ifx\svgscale\undefined%
      \relax%
    \else%
      \setlength{\unitlength}{\unitlength * \real{\svgscale}}%
    \fi%
  \else%
    \setlength{\unitlength}{\svgwidth}%
  \fi%
  \global\let\svgwidth\undefined%
  \global\let\svgscale\undefined%
  \makeatother%
  \begin{picture}(1,0.41214266)%
    \lineheight{1}%
    \setlength\tabcolsep{0pt}%
    \put(0,0){\includegraphics[width=\unitlength,page=1]{ChargeAccNoRadFig2.pdf}}%
    \put(0.20621229,0.30863269){\color[rgb]{0,0,0}\makebox(0,0)[lt]{\lineheight{1.25}\smash{\begin{tabular}[t]{l}$d$\end{tabular}}}}%
    \put(0.91798647,0.07010641){\color[rgb]{0,0,0}\makebox(0,0)[lt]{\lineheight{1.25}\smash{\begin{tabular}[t]{l}$x$\end{tabular}}}}%
    \put(0.34598066,0.06683262){\color[rgb]{0,0,0}\makebox(0,0)[lt]{\lineheight{1.25}\smash{\begin{tabular}[t]{l}$-a$\end{tabular}}}}%
    \put(0.1978575,0.24208737){\color[rgb]{0,0,0}\makebox(0,0)[lt]{\lineheight{1.25}\smash{\begin{tabular}[t]{l}$\Psi_L$\end{tabular}}}}%
    \put(0.66186811,0.12100845){\color[rgb]{0,0,0}\makebox(0,0)[lt]{\lineheight{1.25}\smash{\begin{tabular}[t]{l}${\bf A}$\end{tabular}}}}%
    \put(0.42195942,0.37394386){\color[rgb]{0,0,0}\makebox(0,0)[lt]{\lineheight{1.25}\smash{\begin{tabular}[t]{l}$z$\end{tabular}}}}%
    \put(0.40376344,0.25906871){\color[rgb]{0,0,0}\makebox(0,0)[lt]{\lineheight{1.25}\smash{\begin{tabular}[t]{l}$r$\end{tabular}}}}%
    \put(0.39070319,0.16963293){\color[rgb]{0,0,0}\makebox(0,0)[lt]{\lineheight{1.25}\smash{\begin{tabular}[t]{l}$z_0$\end{tabular}}}}%
    \put(0,0){\includegraphics[width=\unitlength,page=2]{ChargeAccNoRadFig2.pdf}}%
    \put(0.44884863,0.09433692){\color[rgb]{0,0,0}\makebox(0,0)[lt]{\lineheight{1.25}\smash{\begin{tabular}[t]{l}$\Phi$\end{tabular}}}}%
  \end{picture}%
\endgroup%

%% file: ChargeAccNoRadFig3.pdf_tex
\begingroup%
  \makeatletter%
  \providecommand\color[2][]{%
    \errmessage{(Inkscape) Color is used for the text in Inkscape, but the package 'color.sty' is not loaded}%
    \renewcommand\color[2][]{}%
  }%
  \providecommand\transparent[1]{%
    \errmessage{(Inkscape) Transparency is used (non-zero) for the text in Inkscape, but the package 'transparent.sty' is not loaded}%
    \renewcommand\transparent[1]{}%
  }%
  \providecommand\rotatebox[2]{#2}%
  \newcommand*\fsize{\dimexpr\f@size pt\relax}%
  \newcommand*\lineheight[1]{\fontsize{\fsize}{#1\fsize}\selectfont}%
  \ifx\svgwidth\undefined%
    \setlength{\unitlength}{346.8787342bp}%
    \ifx\svgscale\undefined%
      \relax%
    \else%
      \setlength{\unitlength}{\unitlength * \real{\svgscale}}%
    \fi%
  \else%
    \setlength{\unitlength}{\svgwidth}%
  \fi%
  \global\let\svgwidth\undefined%
  \global\let\svgscale\undefined%
  \makeatother%
  \begin{picture}(1,0.41070189)%
    \lineheight{1}%
    \setlength\tabcolsep{0pt}%
    \put(0,0){\includegraphics[width=\unitlength,page=1]{ChargeAccNoRadFig3.pdf}}%
    \put(0.68096688,0.30526393){\color[rgb]{0,0,0}\makebox(0,0)[lt]{\lineheight{1.25}\smash{\begin{tabular}[t]{l}$d$\end{tabular}}}}%
    \put(0.9079669,0.07239887){\color[rgb]{0,0,0}\makebox(0,0)[lt]{\lineheight{1.25}\smash{\begin{tabular}[t]{l}$x$\end{tabular}}}}%
    \put(0.51725923,0.07298293){\color[rgb]{0,0,0}\makebox(0,0)[lt]{\lineheight{1.25}\smash{\begin{tabular}[t]{l}$a$\end{tabular}}}}%
    \put(0.6727031,0.23944496){\color[rgb]{0,0,0}\makebox(0,0)[lt]{\lineheight{1.25}\smash{\begin{tabular}[t]{l}$\Psi_R$\end{tabular}}}}%
    \put(0.78001079,0.12733194){\color[rgb]{0,0,0}\makebox(0,0)[lt]{\lineheight{1.25}\smash{\begin{tabular}[t]{l}${\bf A}$\end{tabular}}}}%
    \put(0.41735385,0.37292002){\color[rgb]{0,0,0}\makebox(0,0)[lt]{\lineheight{1.25}\smash{\begin{tabular}[t]{l}$z$\end{tabular}}}}%
    \put(0.87636182,0.25624094){\color[rgb]{0,0,0}\makebox(0,0)[lt]{\lineheight{1.25}\smash{\begin{tabular}[t]{l}$r$\end{tabular}}}}%
    \put(0.53015188,0.16243032){\color[rgb]{0,0,0}\makebox(0,0)[lt]{\lineheight{1.25}\smash{\begin{tabular}[t]{l}$z_0$\end{tabular}}}}%
    \put(0,0){\includegraphics[width=\unitlength,page=2]{ChargeAccNoRadFig3.pdf}}%
    \put(0.44394957,0.09636491){\color[rgb]{0,0,0}\makebox(0,0)[lt]{\lineheight{1.25}\smash{\begin{tabular}[t]{l}$\Phi$\end{tabular}}}}%
  \end{picture}%
\endgroup%

%% file: ChargeAccNoRadFig4.pdf_tex
\begingroup%
  \makeatletter%
  \providecommand\color[2][]{%
    \errmessage{(Inkscape) Color is used for the text in Inkscape, but the package 'color.sty' is not loaded}%
    \renewcommand\color[2][]{}%
  }%
  \providecommand\transparent[1]{%
    \errmessage{(Inkscape) Transparency is used (non-zero) for the text in Inkscape, but the package 'transparent.sty' is not loaded}%
    \renewcommand\transparent[1]{}%
  }%
  \providecommand\rotatebox[2]{#2}%
  \newcommand*\fsize{\dimexpr\f@size pt\relax}%
  \newcommand*\lineheight[1]{\fontsize{\fsize}{#1\fsize}\selectfont}%
  \ifx\svgwidth\undefined%
    \setlength{\unitlength}{346.8787342bp}%
    \ifx\svgscale\undefined%
      \relax%
    \else%
      \setlength{\unitlength}{\unitlength * \real{\svgscale}}%
    \fi%
  \else%
    \setlength{\unitlength}{\svgwidth}%
  \fi%
  \global\let\svgwidth\undefined%
  \global\let\svgscale\undefined%
  \makeatother%
  \begin{picture}(1,0.41070189)%
    \lineheight{1}%
    \setlength\tabcolsep{0pt}%
    \put(0,0){\includegraphics[width=\unitlength,page=1]{ChargeAccNoRadFig4.pdf}}%
    \put(0.34787213,0.07338303){\color[rgb]{0,0,0}\makebox(0,0)[lt]{\lineheight{1.25}\smash{\begin{tabular}[t]{l}$-a$\end{tabular}}}}%
    \put(0,0){\includegraphics[width=\unitlength,page=2]{ChargeAccNoRadFig4.pdf}}%
    \put(0.68096689,0.30526403){\color[rgb]{0,0,0}\makebox(0,0)[lt]{\lineheight{1.25}\smash{\begin{tabular}[t]{l}$d$\end{tabular}}}}%
    \put(0.9079669,0.07239897){\color[rgb]{0,0,0}\makebox(0,0)[lt]{\lineheight{1.25}\smash{\begin{tabular}[t]{l}$x$\end{tabular}}}}%
    \put(0.51725923,0.07298303){\color[rgb]{0,0,0}\makebox(0,0)[lt]{\lineheight{1.25}\smash{\begin{tabular}[t]{l}$a$\end{tabular}}}}%
    \put(0.6727031,0.23944506){\color[rgb]{0,0,0}\makebox(0,0)[lt]{\lineheight{1.25}\smash{\begin{tabular}[t]{l}$\Psi_R$\end{tabular}}}}%
    \put(0.78001079,0.12733204){\color[rgb]{0,0,0}\makebox(0,0)[lt]{\lineheight{1.25}\smash{\begin{tabular}[t]{l}${\bf A}$\end{tabular}}}}%
    \put(0.41735385,0.37292012){\color[rgb]{0,0,0}\makebox(0,0)[lt]{\lineheight{1.25}\smash{\begin{tabular}[t]{l}$z$\end{tabular}}}}%
    \put(0.87636183,0.25624104){\color[rgb]{0,0,0}\makebox(0,0)[lt]{\lineheight{1.25}\smash{\begin{tabular}[t]{l}$r$\end{tabular}}}}%
    \put(0.53015188,0.16243043){\color[rgb]{0,0,0}\makebox(0,0)[lt]{\lineheight{1.25}\smash{\begin{tabular}[t]{l}$z_0$\end{tabular}}}}%
    \put(0,0){\includegraphics[width=\unitlength,page=3]{ChargeAccNoRadFig4.pdf}}%
    \put(0.40224751,0.25681157){\color[rgb]{0,0,0}\makebox(0,0)[lt]{\lineheight{1.25}\smash{\begin{tabular}[t]{l}$r$\end{tabular}}}}%
    \put(0.21009579,0.30583466){\color[rgb]{0,0,0}\makebox(0,0)[lt]{\lineheight{1.25}\smash{\begin{tabular}[t]{l}$d$\end{tabular}}}}%
    \put(0,0){\includegraphics[width=\unitlength,page=4]{ChargeAccNoRadFig4.pdf}}%
    \put(0.3936541,0.16294696){\color[rgb]{0,0,0}\makebox(0,0)[lt]{\lineheight{1.25}\smash{\begin{tabular}[t]{l}$z_0$\end{tabular}}}}%
    \put(0,0){\includegraphics[width=\unitlength,page=5]{ChargeAccNoRadFig4.pdf}}%
    \put(0.44394958,0.09636501){\color[rgb]{0,0,0}\makebox(0,0)[lt]{\lineheight{1.25}\smash{\begin{tabular}[t]{l}$\Phi$\end{tabular}}}}%
    \put(0.2030274,0.23977537){\color[rgb]{0,0,0}\makebox(0,0)[lt]{\lineheight{1.25}\smash{\begin{tabular}[t]{l}$\Psi_L$\end{tabular}}}}%
    \put(0,0){\includegraphics[width=\unitlength,page=6]{ChargeAccNoRadFig4.pdf}}%
  \end{picture}%
\endgroup%

%% file: ChargeAccNoRadFig6.pdf_tex
\begingroup%
  \makeatletter%
  \providecommand\color[2][]{%
    \errmessage{(Inkscape) Color is used for the text in Inkscape, but the package 'color.sty' is not loaded}%
    \renewcommand\color[2][]{}%
  }%
  \providecommand\transparent[1]{%
    \errmessage{(Inkscape) Transparency is used (non-zero) for the text in Inkscape, but the package 'transparent.sty' is not loaded}%
    \renewcommand\transparent[1]{}%
  }%
  \providecommand\rotatebox[2]{#2}%
  \newcommand*\fsize{\dimexpr\f@size pt\relax}%
  \newcommand*\lineheight[1]{\fontsize{\fsize}{#1\fsize}\selectfont}%
  \ifx\svgwidth\undefined%
    \setlength{\unitlength}{359.64867539bp}%
    \ifx\svgscale\undefined%
      \relax%
    \else%
      \setlength{\unitlength}{\unitlength * \real{\svgscale}}%
    \fi%
  \else%
    \setlength{\unitlength}{\svgwidth}%
  \fi%
  \global\let\svgwidth\undefined%
  \global\let\svgscale\undefined%
  \makeatother%
  \begin{picture}(1,0.50364214)%
    \lineheight{1}%
    \setlength\tabcolsep{0pt}%
    \put(0.51097169,0.14045321){\color[rgb]{0,0,0}\makebox(0,0)[lt]{\lineheight{1.25}\smash{\begin{tabular}[t]{l}+\end{tabular}}}}%
    \put(0.51097169,0.03520169){\color[rgb]{0,0,0}\makebox(0,0)[lt]{\lineheight{1.25}\smash{\begin{tabular}[t]{l}+\end{tabular}}}}%
    \put(0.51097169,0.24570473){\color[rgb]{0,0,0}\makebox(0,0)[lt]{\lineheight{1.25}\smash{\begin{tabular}[t]{l}+\end{tabular}}}}%
    \put(0,0){\includegraphics[width=\unitlength,page=1]{ChargeAccNoRadFig6.pdf}}%
    \put(0.92176207,0.07194752){\color[rgb]{0,0,0}\makebox(0,0)[lt]{\lineheight{1.25}\smash{\begin{tabular}[t]{l}$x$\end{tabular}}}}%
    \put(0.69780089,0.19767247){\color[rgb]{0,0,0}\makebox(0,0)[lt]{\lineheight{1.25}\smash{\begin{tabular}[t]{l}$\Psi_R$\end{tabular}}}}%
    \put(-0.00265185,0.45174739){\color[rgb]{0,0,0}\makebox(0,0)[lt]{\lineheight{1.25}\smash{\begin{tabular}[t]{l}$z$\end{tabular}}}}%
    \put(0,0){\includegraphics[width=\unitlength,page=2]{ChargeAccNoRadFig6.pdf}}%
    \put(0.24037805,0.19946565){\color[rgb]{0,0,0}\makebox(0,0)[lt]{\lineheight{1.25}\smash{\begin{tabular}[t]{l}$\Psi_L$\end{tabular}}}}%
    \put(0,0){\includegraphics[width=\unitlength,page=3]{ChargeAccNoRadFig6.pdf}}%
    \put(0.50526194,0.29845684){\color[rgb]{0,0,0}\makebox(0,0)[lt]{\lineheight{1.25}\smash{\begin{tabular}[t]{l}$C$\end{tabular}}}}%
    \put(0,0){\includegraphics[width=\unitlength,page=4]{ChargeAccNoRadFig6.pdf}}%
    \put(0.51083578,0.35068471){\color[rgb]{0,0,0}\makebox(0,0)[lt]{\lineheight{1.25}\smash{\begin{tabular}[t]{l}+\end{tabular}}}}%
    \put(0.50047364,0.47738304){\color[rgb]{0,0,0}\makebox(0,0)[lt]{\lineheight{1.25}\smash{\begin{tabular}[t]{l}+\end{tabular}}}}%
    \put(0.45305749,0.47936159){\color[rgb]{0,0,0}\makebox(0,0)[lt]{\lineheight{1.25}\smash{\begin{tabular}[t]{l}-\end{tabular}}}}%
    \put(0.44071613,0.35424802){\color[rgb]{0,0,0}\makebox(0,0)[lt]{\lineheight{1.25}\smash{\begin{tabular}[t]{l}-\end{tabular}}}}%
    \put(0.44071606,0.03918402){\color[rgb]{0,0,0}\makebox(0,0)[lt]{\lineheight{1.25}\smash{\begin{tabular}[t]{l}-\end{tabular}}}}%
    \put(0.44071606,0.24922667){\color[rgb]{0,0,0}\makebox(0,0)[lt]{\lineheight{1.25}\smash{\begin{tabular}[t]{l}-\end{tabular}}}}%
    \put(0.44071606,0.14420527){\color[rgb]{0,0,0}\makebox(0,0)[lt]{\lineheight{1.25}\smash{\begin{tabular}[t]{l}-\end{tabular}}}}%
  \end{picture}%
\endgroup%

%% file: ChargeAccNoRadFig5.pdf_tex
\begingroup%
  \makeatletter%
  \providecommand\color[2][]{%
    \errmessage{(Inkscape) Color is used for the text in Inkscape, but the package 'color.sty' is not loaded}%
    \renewcommand\color[2][]{}%
  }%
  \providecommand\transparent[1]{%
    \errmessage{(Inkscape) Transparency is used (non-zero) for the text in Inkscape, but the package 'transparent.sty' is not loaded}%
    \renewcommand\transparent[1]{}%
  }%
  \providecommand\rotatebox[2]{#2}%
  \newcommand*\fsize{\dimexpr\f@size pt\relax}%
  \newcommand*\lineheight[1]{\fontsize{\fsize}{#1\fsize}\selectfont}%
  \ifx\svgwidth\undefined%
    \setlength{\unitlength}{396.66444793bp}%
    \ifx\svgscale\undefined%
      \relax%
    \else%
      \setlength{\unitlength}{\unitlength * \real{\svgscale}}%
    \fi%
  \else%
    \setlength{\unitlength}{\svgwidth}%
  \fi%
  \global\let\svgwidth\undefined%
  \global\let\svgscale\undefined%
  \makeatother%
  \begin{picture}(1,0.35222863)%
    \lineheight{1}%
    \setlength\tabcolsep{0pt}%
    \put(0,0){\includegraphics[width=\unitlength,page=1]{ChargeAccNoRadFig5.pdf}}%
    \put(0.80762089,0.05976324){\color[rgb]{0,0,0}\makebox(0,0)[lt]{\lineheight{1.25}\smash{\begin{tabular}[t]{l}$x$\end{tabular}}}}%
    \put(0.7661418,0.1125958){\color[rgb]{0,0,0}\makebox(0,0)[lt]{\lineheight{1.25}\smash{\begin{tabular}[t]{l}${\bf A}$\end{tabular}}}}%
    \put(-0.00240438,0.31500261){\color[rgb]{0,0,0}\makebox(0,0)[lt]{\lineheight{1.25}\smash{\begin{tabular}[t]{l}$z$\end{tabular}}}}%
    \put(0.26079058,0.10316965){\color[rgb]{0,0,0}\makebox(0,0)[lt]{\lineheight{1.25}\smash{\begin{tabular}[t]{l}$\alpha$\end{tabular}}}}%
    \put(0.46499339,0.10946283){\color[rgb]{0,0,0}\makebox(0,0)[lt]{\lineheight{1.25}\smash{\begin{tabular}[t]{l}$2 \alpha$\end{tabular}}}}%
    \put(0,0){\includegraphics[width=\unitlength,page=2]{ChargeAccNoRadFig5.pdf}}%
    \put(0.10658339,0.26484561){\color[rgb]{0,0,0}\makebox(0,0)[lt]{\lineheight{1.25}\smash{\begin{tabular}[t]{l}$d$\end{tabular}}}}%
    \put(0,0){\includegraphics[width=\unitlength,page=3]{ChargeAccNoRadFig5.pdf}}%
    \put(0.36901584,0.30553929){\color[rgb]{0,0,0}\makebox(0,0)[lt]{\lineheight{1.25}\smash{\begin{tabular}[t]{l}$L$\end{tabular}}}}%
    \put(0,0){\includegraphics[width=\unitlength,page=4]{ChargeAccNoRadFig5.pdf}}%
    \put(0.12661151,0.17797968){\color[rgb]{0,0,0}\makebox(0,0)[lt]{\lineheight{1.25}\smash{\begin{tabular}[t]{l}$l$\end{tabular}}}}%
    \put(0,0){\includegraphics[width=\unitlength,page=5]{ChargeAccNoRadFig5.pdf}}%
    \put(0.20808812,0.08194346){\color[rgb]{0,0,0}\makebox(0,0)[lt]{\lineheight{1.25}\smash{\begin{tabular}[t]{l}$\Phi$\end{tabular}}}}%
    \put(0.06825913,0.22125816){\color[rgb]{0,0,0}\makebox(0,0)[lt]{\lineheight{1.25}\smash{\begin{tabular}[t]{l}$\Theta(x|d)$\end{tabular}}}}%
    \put(0,0){\includegraphics[width=\unitlength,page=6]{ChargeAccNoRadFig5.pdf}}%
    \put(0.60491014,0.11748472){\color[rgb]{0,0,0}\makebox(0,0)[lt]{\lineheight{1.25}\smash{\begin{tabular}[t]{l}$(N-1) \alpha$\end{tabular}}}}%
    \put(0.41451959,0.0819857){\color[rgb]{0,0,0}\makebox(0,0)[lt]{\lineheight{1.25}\smash{\begin{tabular}[t]{l}$\Phi$\end{tabular}}}}%
    \put(0,0){\includegraphics[width=\unitlength,page=7]{ChargeAccNoRadFig5.pdf}}%
    \put(0.56310682,0.08149301){\color[rgb]{0,0,0}\makebox(0,0)[lt]{\lineheight{1.25}\smash{\begin{tabular}[t]{l}$\Phi$\end{tabular}}}}%
    \put(0.24254758,0.22035505){\color[rgb]{0,0,0}\makebox(0,0)[lt]{\lineheight{1.25}\smash{\begin{tabular}[t]{l}$\Theta(x-l|d)$\end{tabular}}}}%
    \put(0,0){\includegraphics[width=\unitlength,page=8]{ChargeAccNoRadFig5.pdf}}%
    \put(0.59944371,0.22140972){\color[rgb]{0,0,0}\makebox(0,0)[lt]{\lineheight{1.25}\smash{\begin{tabular}[t]{l}$\Theta(x-(N-1)l|d)$\end{tabular}}}}%
    \put(0,0){\includegraphics[width=\unitlength,page=9]{ChargeAccNoRadFig5.pdf}}%
    \put(0.44841043,0.22046426){\color[rgb]{0,0,0}\makebox(0,0)[lt]{\lineheight{1.25}\smash{\begin{tabular}[t]{l}$\Theta(x-$\end{tabular}}}}%
    \put(0,0){\includegraphics[width=\unitlength,page=10]{ChargeAccNoRadFig5.pdf}}%
  \end{picture}%
\endgroup%

%% file: ChargeAccNoRad.bbl
\begin{thebibliography}{7}%
\makeatletter
\providecommand \@ifxundefined [1]{%
 \@ifx{#1\undefined}
}%
\providecommand \@ifnum [1]{%
 \ifnum #1\expandafter \@firstoftwo
 \else \expandafter \@secondoftwo
 \fi
}%
\providecommand \@ifx [1]{%
 \ifx #1\expandafter \@firstoftwo
 \else \expandafter \@secondoftwo
 \fi
}%
\providecommand \natexlab [1]{#1}%
\providecommand \enquote  [1]{``#1''}%
\providecommand \bibnamefont  [1]{#1}%
\providecommand \bibfnamefont [1]{#1}%
\providecommand \citenamefont [1]{#1}%
\providecommand \href@noop [0]{\@secondoftwo}%
\providecommand \href [0]{\begingroup \@sanitize@url \@href}%
\providecommand \@href[1]{\@@startlink{#1}\@@href}%
\providecommand \@@href[1]{\endgroup#1\@@endlink}%
\providecommand \@sanitize@url [0]{\catcode `\\12\catcode `\$12\catcode `\&12\catcode `\#12\catcode `\^12\catcode `\_12\catcode `\%12\relax}%
\providecommand \@@startlink[1]{}%
\providecommand \@@endlink[0]{}%
\providecommand \url  [0]{\begingroup\@sanitize@url \@url }%
\providecommand \@url [1]{\endgroup\@href {#1}{\urlprefix }}%
\providecommand \urlprefix  [0]{URL }%
\providecommand \Eprint [0]{\href }%
\providecommand \doibase [0]{https://doi.org/}%
\providecommand \selectlanguage [0]{\@gobble}%
\providecommand \bibinfo  [0]{\@secondoftwo}%
\providecommand \bibfield  [0]{\@secondoftwo}%
\providecommand \translation [1]{[#1]}%
\providecommand \BibitemOpen [0]{}%
\providecommand \bibitemStop [0]{}%
\providecommand \bibitemNoStop [0]{.\EOS\space}%
\providecommand \EOS [0]{\spacefactor3000\relax}%
\providecommand \BibitemShut  [1]{\csname bibitem#1\endcsname}%
\let\auto@bib@innerbib\@empty
\bibitem [{\citenamefont {Maxwell}(1865)}]{Maxwellemwaves}%
  \BibitemOpen
  \bibfield  {author} {\bibinfo {author} {\bibfnamefont {J.~C.}\ \bibnamefont {Maxwell}},\ }\bibfield  {title} {\bibinfo {title} {A dynamical theory of the electromagnetic field},\ }\href {https://doi.org/10.1098/rstl.1865.0008} {\bibfield  {journal} {\bibinfo  {journal} {Philos. Trans. R. Soc. London}\ ,\ \bibinfo {pages} {459}} (\bibinfo {year} {1865})}\BibitemShut {NoStop}%
\bibitem [{\citenamefont {Aharonov}\ and\ \citenamefont {Bohm}(1959)}]{aharonovBohm}%
  \BibitemOpen
  \bibfield  {author} {\bibinfo {author} {\bibfnamefont {Y.}~\bibnamefont {Aharonov}}\ and\ \bibinfo {author} {\bibfnamefont {D.}~\bibnamefont {Bohm}},\ }\bibfield  {title} {\bibinfo {title} {Significance of electromagnetic potentials in the quantum theory},\ }\href {https://doi.org/10.1103/PhysRev.115.485} {\bibfield  {journal} {\bibinfo  {journal} {Phys. Rev.}\ }\textbf {\bibinfo {volume} {115}},\ \bibinfo {pages} {485} (\bibinfo {year} {1959})}\BibitemShut {NoStop}%
\bibitem [{\citenamefont {Aharonov}\ \emph {et~al.}(1969)\citenamefont {Aharonov}, \citenamefont {Pendleton},\ and\ \citenamefont {Petersen}}]{modularVariables}%
  \BibitemOpen
  \bibfield  {author} {\bibinfo {author} {\bibfnamefont {Y.}~\bibnamefont {Aharonov}}, \bibinfo {author} {\bibfnamefont {H.}~\bibnamefont {Pendleton}},\ and\ \bibinfo {author} {\bibfnamefont {A.}~\bibnamefont {Petersen}},\ }\bibfield  {title} {\bibinfo {title} {Modular variables in quantum theory},\ }\href {https://doi.org/10.1007/BF00670008} {\bibfield  {journal} {\bibinfo  {journal} {Int. J. Theor. Phys.}\ }\textbf {\bibinfo {volume} {2}},\ \bibinfo {pages} {213} (\bibinfo {year} {1969})}\BibitemShut {NoStop}%
\bibitem [{\citenamefont {Aharonov}(1984)}]{yakirModularMomentum}%
  \BibitemOpen
  \bibfield  {author} {\bibinfo {author} {\bibfnamefont {Y.}~\bibnamefont {Aharonov}},\ }\bibfield  {title} {\bibinfo {title} {Nonlocal phenomena and the {A}haronov-{B}ohm effect},\ }in\ \href {https://doi.org/10.1142/9789812819895_0002} {\emph {\bibinfo {booktitle} {Foundations of Quantum Mechanics in Light of New Technology [Proceedings of the International Symposium, Tokyo, August 1983]}}},\ \bibinfo {series} {Physical Society of Japan, Tokyo, Japan}, Vol.\ \bibinfo {volume} {1509},\ \bibinfo {editor} {edited by\ \bibinfo {editor} {\bibfnamefont {S.}~\bibnamefont {Kamefuchi~et al.}}}\ (\bibinfo {year} {1984})\ pp.\ \bibinfo {pages} {10--19}\BibitemShut {NoStop}%
\bibitem [{\citenamefont {Popescu}(2010)}]{dynamicalNonlocality}%
  \BibitemOpen
  \bibfield  {author} {\bibinfo {author} {\bibfnamefont {S.}~\bibnamefont {Popescu}},\ }\bibfield  {title} {\bibinfo {title} {Dynamical quantum non-locality},\ }\href {https://doi.org/10.1038/nphys1619} {\bibfield  {journal} {\bibinfo  {journal} {Nat. Phys.}\ }\textbf {\bibinfo {volume} {6}},\ \bibinfo {pages} {151} (\bibinfo {year} {2010})}\BibitemShut {NoStop}%
\bibitem [{\citenamefont {Thomson}(1904)}]{thomson}%
  \BibitemOpen
  \bibfield  {author} {\bibinfo {author} {\bibfnamefont {J.~J.}\ \bibnamefont {Thomson}},\ }\href@noop {} {\emph {\bibinfo {title} {Electricity and Matter}}}\ (\bibinfo  {publisher} {Archibald Constable},\ \bibinfo {address} {London, UK},\ \bibinfo {year} {1904})\ Chap.~\bibinfo {chapter} {3}, pp.\ \bibinfo {pages} {59--61}\BibitemShut {NoStop}%
\bibitem [{\citenamefont {Feynman}\ \emph {et~al.}(1964)\citenamefont {Feynman}, \citenamefont {Leighton},\ and\ \citenamefont {Sands}}]{feynmanI}%
  \BibitemOpen
  \bibfield  {author} {\bibinfo {author} {\bibfnamefont {R.~P.}\ \bibnamefont {Feynman}}, \bibinfo {author} {\bibfnamefont {R.~B.}\ \bibnamefont {Leighton}},\ and\ \bibinfo {author} {\bibfnamefont {M.}~\bibnamefont {Sands}},\ }\href {https://www.feynmanlectures.caltech.edu} {\emph {\bibinfo {title} {The Feynman Lectures on Physics}}},\ Vol.~\bibinfo {volume} {I}\ (\bibinfo  {publisher} {Addison-Wesley Publishing Company, Inc.},\ \bibinfo {address} {Reading, Massachusetts},\ \bibinfo {year} {1964})\ Chap.\ \bibinfo {chapter} {28-1, 28-2}\BibitemShut {NoStop}%
\end{thebibliography}%
